\documentstyle[preprint,aps,epsf,floats]{revtex}    
\def\bq{\begin{equation}}
\def\eq{\end{equation}}
\def\ba{\begin{eqnarray}}
\def\ea{\end{eqnarray}}

\begin{document}
\thispagestyle{empty}
  
\renewcommand{\small}{\normalsize} 

\preprint{
\font\fortssbx=cmssbx10 scaled \magstep2
\hbox to \hsize{
\hskip.5in \raise.1in\hbox{\fortssbx University of Wisconsin - Madison} 
\hfill\vtop{\hbox{\bf MADPH-97-1023}
            \hbox{December 1997}} }
}
  
\title{\vspace{.5in}
Searching for $H \rightarrow \gamma \gamma$ in weak boson fusion
at the LHC
}

\author{D.~Rainwater and D.~Zeppenfeld}
\address{
Department of Physics, University of Wisconsin, Madison, WI 53706 
}
\maketitle
\begin{abstract}
Weak boson fusion is a copious source of intermediate mass Higgs bosons at 
the LHC, with a rate $\sigma B(H \rightarrow \gamma \gamma)$ of up to 9~fb.
The additional very energetic forward jets in these events provide for
a unique signature. A parton level analysis of the dominant backgrounds 
demonstrates that this channel allows the observation of 
$H \rightarrow \gamma \gamma$ in a low background environment, with 
modest luminosity.
\end{abstract}
%
%

\newpage

%
%

The search for the Higgs boson and, hence, for the source of electroweak 
symmetry breaking and fermion mass generation, 
remains one of the premier tasks of present and future high energy physics 
experiments. Fits to precision electroweak (EW) data have for 
some time suggested a relatively low Higgs boson mass, in the 100~GeV 
range~\cite{EWfits} and this is one of the reasons why the search for 
an intermediate mass Higgs boson is particularly important~\cite{Dawson}.  
Beyond the reach of LEP, for masses in the $100-150$~GeV range,
the $H\to \gamma\gamma$ decay channel at the CERN LHC is very promising.
Consequently, LHC detectors are designed with excellent photon detection
capabilities, resulting in a di-photon mass resolution of order 1~GeV for
a Higgs boson mass around 120~GeV~\cite{CMS-ATLAS}. Another advantage of the
$H\to \gamma\gamma$ channel, in particular compared to the dominant
$H\to b\bar b$ mode, is the lower background from QCD processes. 

For this intermediate mass range, most of the literature has focussed on
Higgs production via gluon fusion~\cite{Dawson},
and $t\bar{t}H$~\cite{ttH} or $WH(ZH)$~\cite{WH} associated production.
While production via gluon fusion has the largest cross section by about an
order of magnitude, there are substantial QCD backgrounds but
few handles to distinguish them from the signal.  Essentially, only the decay
photons' transverse momentum and the sharp resonance in the $\gamma\gamma$
invariant mass distribution can be used.

It is necessary to study other production channels for several reasons.  
For instance,  electroweak symmetry breaking and fermion mass generation may
be less intimately connected than in the Standard Model (SM) and the coupling
of the lightest Higgs resonance to fermions might be severely suppressed.
In this case, neither $gg \rightarrow H$ fusion nor $t\bar{t}H$ associated 
production would be observed.  Once the Higgs boson is observed in both 
$gg \rightarrow H$ and the weak boson fusion process $qq \to qqH$, where the 
Higgs is radiated off virtual $W$'s or $Z$'s, the cross section ratio of 
these modes measures the ratio of the Higgs coupling to the top quark
and to W,Z. This value is 
fixed in the SM, but deviations are expected in more general models, like 
supersymmetry with its two Higgs doublets~\cite{susycoupl}.  Finally, as we 
shall demonstrate, the weak boson fusion channel may yield a quicker 
discovery, requiring only 10-20~fb$^{-1}$, which compares favorably to 
the integrated luminosity required for discovery in the 
$gg\to H\to\gamma\gamma$ channel~\cite{CMS-ATLAS}.

\begin{figure}[t]
\vspace*{0.5in}            
\begin{picture}(0,0)(0,0)
\includegraphics{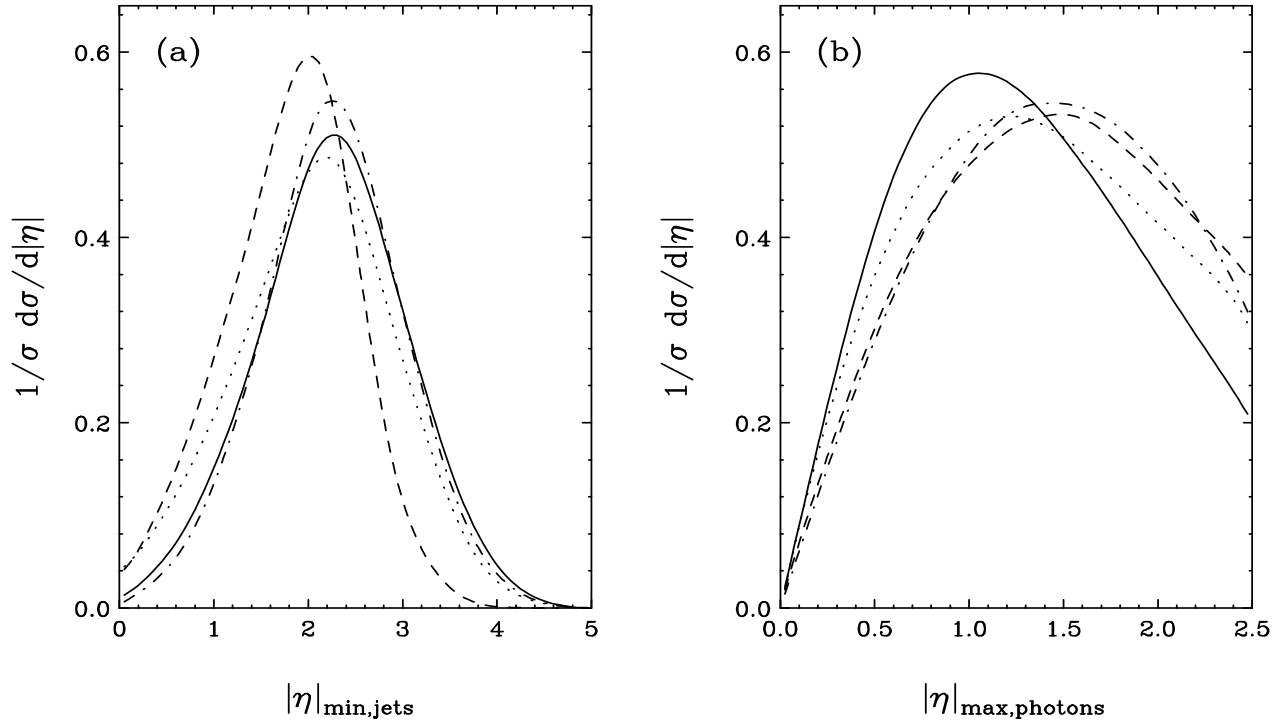}
\end{picture}
\vspace{8.5cm}
\caption{Normalized pseudo-rapidity distributions of (a) the most central 
tagging jet and (b) the photon closest to the beam axis in $jj\gamma\gamma$
events at the LHC. The generic acceptance cuts of Eq.~(\protect\ref{eq:1})
and the forward jet tagging cuts of Eq.~(\protect\ref{eq:2}) are imposed.  
Results are shown for the $qq\to qqH$ signal at $m_H=120$~GeV (solid line) 
for the irreducible QCD background (dashed line), the irreducible EW 
background (dot-dashed line), and for the double parton scattering (DPS)
background (dotted line).
\label{fig:y_dist}
}
\vspace*{0.2in}
\end{figure}

\begin{figure}[t]
\vspace*{0.5in}            
\begin{picture}(0,0)(0,0)
\includegraphics{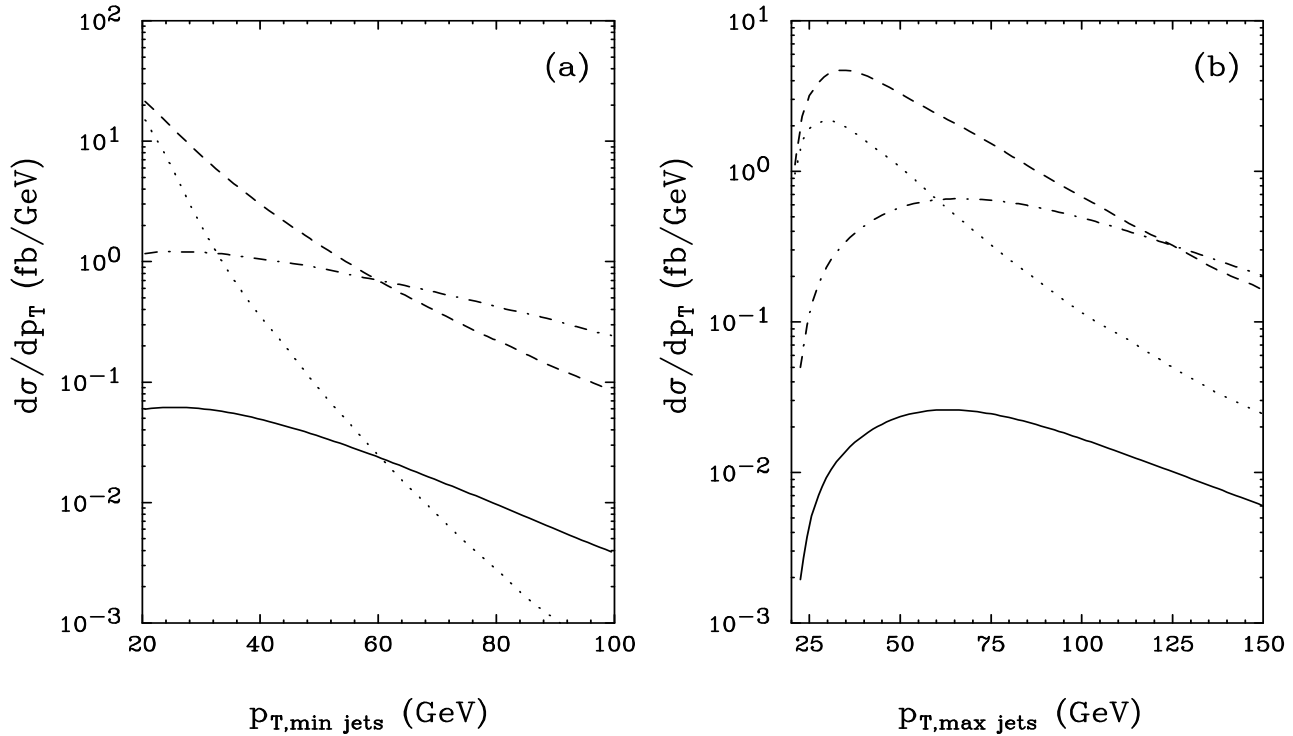}
\end{picture}
\vspace{8.5cm}
\caption{Transverse momentum distributions of (a) the softer and (b) the 
harder of the two tagging jets in $jj\gamma\gamma$ events. Generic acceptance
cuts (Eq.~(\ref{eq:1})) and forward jet tagging cuts (Eq.~(\ref{eq:2})).
are imposed. The signal (solid curve) and the backgrounds are labeled
as in Fig.~\ref{fig:y_dist}.
\label{fig:pT_jets}
}
\vspace*{0.2in}
\end{figure}

\begin{figure}[t]
\vspace*{0.5in}            
\begin{picture}(0,0)(0,0)
\includegraphics{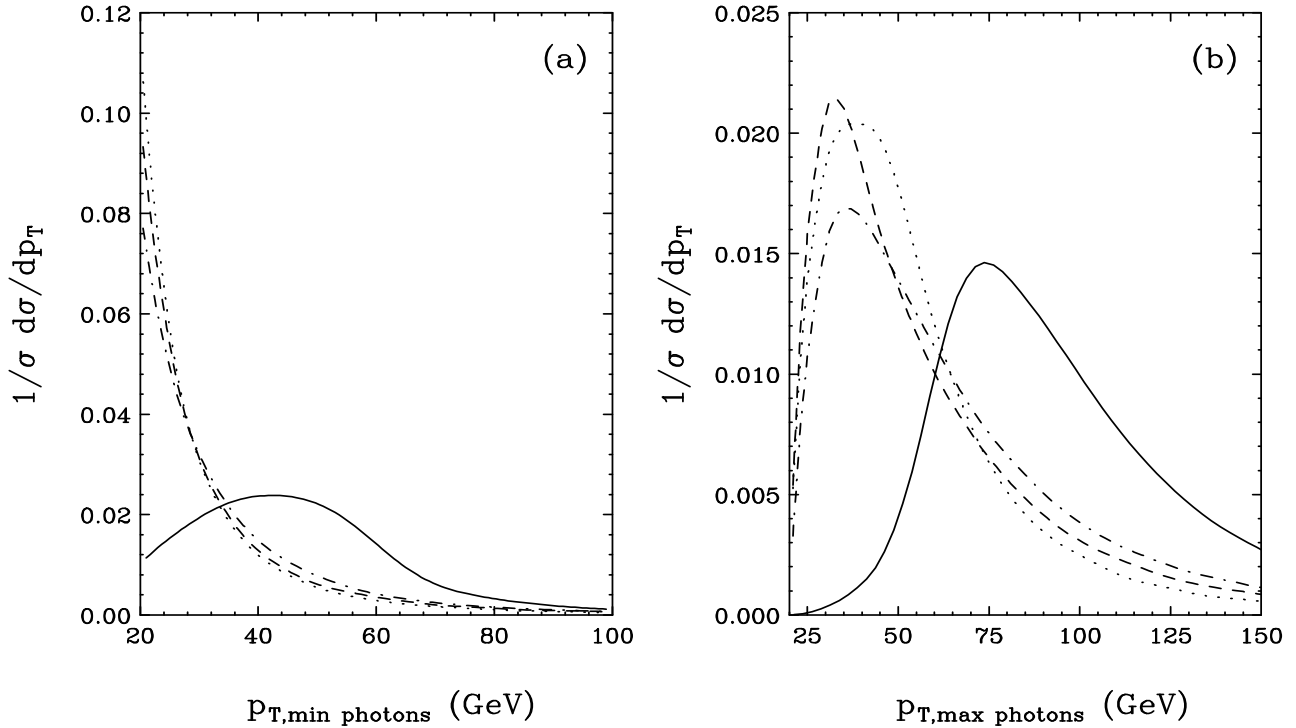}
\end{picture}
\vspace{8.5cm}
\caption{Transverse momentum distributions of (a) the softer and (b) the 
harder of the two photons in $jj\gamma\gamma$ events. Generic acceptance
cuts (Eq.~(\ref{eq:1})) and forward jet tagging cuts (Eq.~(\ref{eq:2}))
are imposed. The signal (solid curve) and the backgrounds are labeled
as in Fig.~\ref{fig:y_dist}.
\label{fig:pT_pho}
}
\vspace*{0.2in}
\end{figure}

Our analysis is a parton-level Monte Carlo study, using full tree-level matrix
elements of the weak boson fusion Higgs signal and the various backgrounds.
Cross sections for Higgs production at the LHC are well-known~\cite{Dawson}.  
For a Higgs in the intermediate
mass range, the weak boson fusion cross section is approximately one order
of magnitude smaller than for gluon fusion.  Features of the signal are a
centrally produced Higgs which tends to yield central photons, and two jets
which enter the detector at large rapidity compared to the photons (see
Fig.~\ref{fig:y_dist}).  Another characteristic feature of the signal 
are the semi-hard transverse momentum distributions of the jets and
photons which are shown in Figs.~\ref{fig:pT_jets} and~\ref{fig:pT_pho}.
The $p_{T_{max}}$ distributions of the signal peak well above
detector thresholds, which allows for higher jet and photon $p_T$ cuts to 
reduce the background while retaining a large signal acceptance. 
For the photons, the growth of the median photon $p_T$  with Higgs mass
improves the signal acceptance when searching at the upper end of the 
intermediate mass range.


The signal can be described, at tree level, by two single-Feynman-diagram 
processes, $WW$ and $ZZ$
fusion where the weak bosons are emitted from the incoming quarks.  For the
$H \rightarrow \gamma\gamma$ partial decay width it is sufficient to include
only the contribution from $t$ and $W$ loops.  As for the backgrounds,
we use CTEQ4M parton distribution functions~\cite{CTEQ4_pdf} and the EW
parameters $m_Z = 91.188$~GeV, $m_t = 175.0$~GeV, $\sin^2 \theta_W = 0.2315$,
and $G_F = 1.16639 \times 10^-5$~GeV$^{-2}$.  We choose the factorization
scale $\mu_f =$ minimum $p_T$ of the jets.

We consider three levels of cuts.  The basic acceptance requirement ensures
that two photons and two jets are observed in the detector, with very high
trigger efficiency~\cite{CMS-ATLAS}:
\bq
\label{eq:1}
\begin{array}{c}
p_{T_j} \geq 20~{\rm GeV}\;, \qquad \; p_{T_{\gamma}} \geq 20~{\rm GeV}\;, \\
|\eta_j| \leq 5.0\;, \qquad \; |\eta_{\gamma}| \leq 2.5\,, \\
\triangle R_{jj} \geq 0.7\;, \qquad \; \triangle R_{j\gamma} \geq 0.7 \, .
\end{array}
\eq
At the second level, double forward jet tagging is required, with two jets
in opposite hemispheres and the photons located between the jets in 
pseudo-rapidity:
\ba
\label{eq:2}
&&\triangle \eta_{tags} = |\eta_{j_1}-\eta_{j_2}| \geq 4.4\;,  \qquad \;
\eta_{j_1} \cdot \eta_{j_2} < 0\; ,  \nonumber \\
&&{\rm min}\{\eta_{j_1},\eta_{j_2}\}+0.7 \: \leq \: \eta_{\gamma} \: \leq \:
{\rm max}\{\eta_{j_1},\eta_{j_2}\}-0.7 \; .
\ea
This technique to separate weak boson scattering from various backgrounds is 
well-established~\cite{Cahn,BCHP,DGOV}, in particular for heavy Higgs boson 
searches. For $m_H = 120$~GeV, the cuts of Eqs.~(\ref{eq:1}) and (\ref{eq:2})
yield cross sections of 5.1 and 2.4~fb, respectively.

Given the features of the signal, we need to consider background processes
that can lead to events with two hard, isolated photons and two forward
jets.  The prime sources are irreducible QCD and EW 
$\: 2\,\gamma + 2\,$jet processes. Double parton scattering (DPS), with pairs
of jets and/or photons arising from two independent partonic collisions in
one $pp$ interaction, is  considered also. We do not consider reducible
backgrounds, where e.g. a jet fragmenting into a leading $\pi^0$ 
is misidentified as a photon.  Reducible backgrounds were shown to be 
small compared to irreducible ones in the analysis of the 
$gg\to H\to\gamma\gamma$ signal~\cite{CMS-ATLAS} and we assume the same to 
hold for the cleaner signal considered here.
Matrix elements for the irreducible QCD processes are available in the
literature~\cite{vvjj}, but we are not aware of previous calculations of 
the irreducible EW background.
To generate the matrix elements for it and the DPS codes we use
Madgraph~\cite{Madgraph}. The running of $\alpha_s$ is determined at leading 
order and we take $\alpha_s(M_Z) = 0.118$ throughout.

The largest background consists of all QCD $2\to 2$ processes which contain 
one or two quark lines, from which the two photons are radiated. Examples
are $q \bar Q\to q\bar Q\gamma\gamma$ or $qg\to qg\gamma\gamma$.
For this irreducible QCD background, the
renormalization scale is chosen as the average $p_T$ of the jets,
$\mu_r = \frac{1}{n_{jet}} \sum p_{T_{jet}}\,$, while the factorization
scale is taken as the average $p_T$ of the jets and photons,
$\mu_f = \frac{1}{n_{part}} \sum p_{T_{all}}\,$.  A prominent feature
of the irreducible QCD background is the steeply falling transverse momentum
distributions of both the jets and photons, as given by the dashed lines in
Figs.~\ref{fig:pT_jets} and~\ref{fig:pT_pho}. These distributions are typical
for bremsstrahlung processes and allow one to suppress the backgrounds further
by harder $p_T$ cuts.  Another feature of the irreducible QCD background
is the generally higher rapidity of the photons (see Fig.~\ref{fig:y_dist}):  
photon bremsstrahlung occurs at small angles with respect to the parent 
quarks, leading to forward photons once the jets are required to be forward.

The irreducible EW background consists of $qQ\to qQ$ processes mediated by
$t$-channel $Z$, $\gamma$, or $W$ exchange, with additional radiation of two 
photons. $\gamma$ and $Z$ exchange processes have amplitudes which are 
proportional to the ones of analogous gluon 
exchange processes, but with smaller couplings. We ignore them because,
in all regions of phase space, they constitute only a tiny correction to 
the irreducible QCD background.  We do include all charged current $qQ\to
qQ\gamma\gamma$ (and crossing related) processes, however. 
$W$ exchange processes can produce central photons by emission from the 
exchanged $W$ and, therefore, are kinematically similar to the
signal.  This signal-like component remains after forward jet tagging cuts,
as can readily be seen in the $p_T$ distribution of the jets in
Fig.~\ref{fig:pT_jets}.  While formally of order $\alpha^4$ and thus
suppressed compared to the order $\alpha^3$ Higgs signal, the small
$H \rightarrow \gamma \gamma$ branching ratio leads to comparable event rates.
Because kinematic cuts on the jets cannot reduce this background compared to
the signal, it is potentially dangerous.
The irreducible EW background is determined with the same choice of 
factorization scale as the irreducible QCD background.

With jet transverse momenta as low as 20~GeV, double parton scattering (DPS) 
is a potential source of backgrounds.  DPS is the
occurrence of two distinct hard scatterings in the collision of a single pair
of protons.  Following Ref.~\cite{DPS_theory}, we calculate the cross 
section for two distinguishable processes, happening in one $pp$ collision, as 
\bq
\label{eq:DPS}
\sigma_{DPS} = \frac{\sigma_{1}\sigma_{2}}{\sigma_{eff}} \;,
\eq
with the additional constraint that the sum of initial parton energies from 
one proton be bounded by the beam energy.  
$\sigma_{eff}$ parameterizes the transverse size of the proton.  It has
recently been measured by CDF as $\sigma_{eff}=14.5$~mb~\cite{CDF_DPS}. 
We assume the same value to hold for LHC energies. 

One DPS background arises from simultaneous $\gamma \gamma j$ and $jj$
events, where the jet in the $\gamma \gamma j$ hard scattering is observed
as a tagging jet, together with one of the two jets in the dijet process.
In order to avoid a three-jet signature, one might want to require the 
second jet of the dijet process to fall outside the acceptance region
of Eq.~(\ref{eq:1}). However, this would severely underestimate this 
DPS contribution, since soft gluon radiation must be taken into account
in a more realistic simulation. Soft radiation destroys the $p_T$ balance 
of the two jets in the dijet process, leading to the possibility of only 
one of the two final state partons to be identified as a jet, even though
both satisfy the pseudo-rapidity requirements of Eq.~(\ref{eq:1}). 
Since our tree-level calculation cannot properly take into account such 
effects, we conservatively select the higher-energy jet of the dijet process
in the hemisphere opposite that of the jet from the $\gamma\gamma j$ event, 
and allow the third jet to be anywhere, completely ignoring it for the 
purposes of imposing further cuts.

A second DPS mode consists of two overlapping $\gamma j$ events.  All final
state particles must be observed above threshold in the detector. With full 
acceptance cuts this background is found to be insignificant compared to the 
others, and will not be considered further. We do not consider
DPS backgrounds from overlapping $\gamma\gamma$ and $jj$ events since
the double forward jet tagging requirements of Eq.~(\ref{eq:2}) force the
dijet invariant mass to be very large, effectively eliminating this 
background.

At the basic level of cuts (Eq.~(\ref{eq:1})), the backgrounds are 
overwhelming, the irreducible QCD component alone being up to two orders of
magnitude larger than the signal in an $m_{\gamma\gamma}=m_H$ invariant mass
bin of width 2~GeV.  This is not surprising:  the presence of $p_T=20$~GeV
jets is a common occurrence in hard scattering events at the LHC.  The double
forward jet tagging requirement of Eq.~(\ref{eq:2}), with its concomitant
large dijet invariant mass, reduces the signal by $\approx$~50\%
but decreases the total background by almost two orders of magnitude, below
the level of the signal.  We can reduce the backgrounds even further by
employing harder $p_T$ cuts on the jets and photons, and find that the
following asymmetric $p_T$ cuts bring the backgrounds down another factor of
three, while accepting over 85\%
of the signal:
\ba
p_{T_{j1}} \geq 40~{\rm GeV}\; , && \qquad p_{T_{j2}} \geq 20~{\rm GeV}\; , 
\nonumber \\ 
\label{eq:3}
p_{T_{\gamma 1}} \geq 50~{\rm GeV}\; , && \qquad 
p_{T_{\gamma 2}} \geq 25~{\rm GeV}\; .
\ea
For $m_H = 120$~GeV, the resulting signal cross section is 2.0~fb.

The effectiveness of these cuts stems from two differing characteristics of
the signal and background.  First, the generic $jj\gamma\gamma$ backgrounds
are produced at small center of mass energies 
and are efficiently suppressed once we require a large invariant mass
for the final state system, via the far forward rapidities of the two
opposite-hemisphere tagging jets.  Second, we expect the Higgs
to be centrally produced, resulting in central photons, while background
photons are primarily from bremsstrahlung off quarks. Since the jets are far
forward, the photons will likewise tend to be at high average $|\eta|$.  In 
addition, bremsstrahlung tends to be soft, and the harder $p_T$ cuts on the 
photons quite efficiently reject bremsstrahlung events.  We could require
even higher $p_{T_{\gamma}}$ cuts, making the backgrounds negligible,
but this would come at the expense of a sizeable reduction in signal rate, 
leaving only a few events in 10 fb$^{-1}$ of data.

\begin{figure}[htb]
\vspace*{0.5in}            
\begin{picture}(0,0)(0,0)
\includegraphics{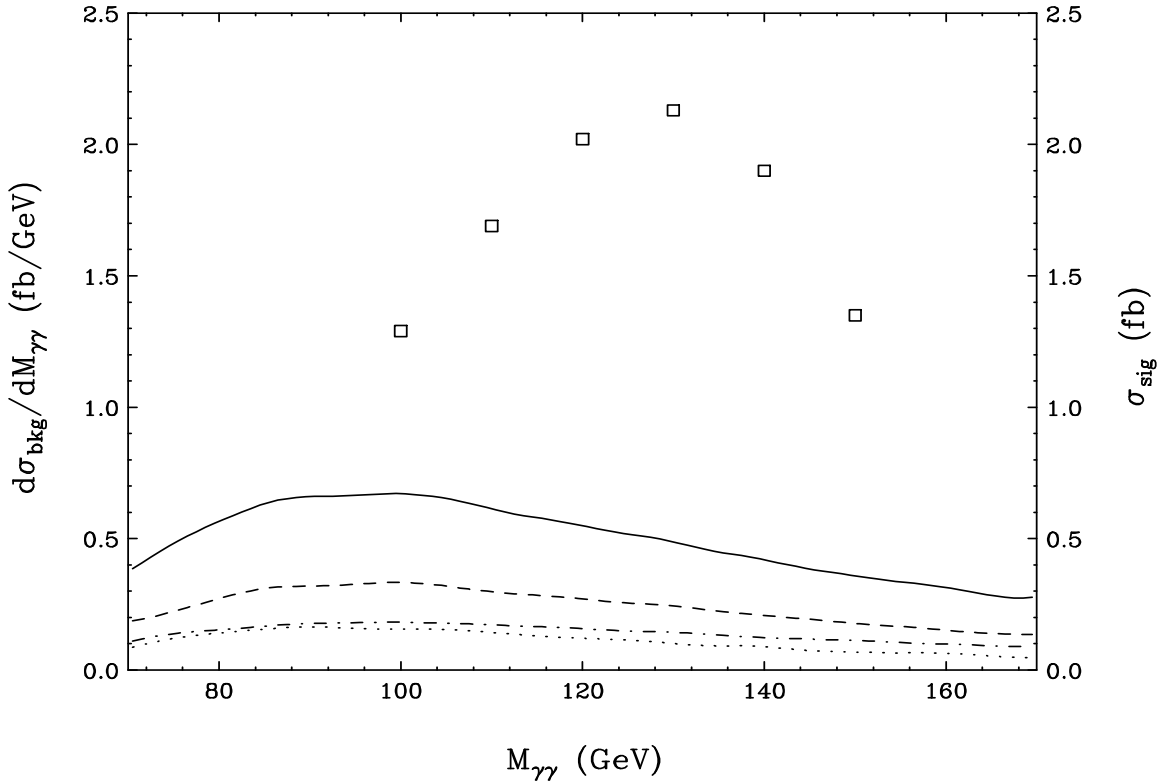}
\end{picture}
\vspace{9.0 cm}
\caption{Higgs boson signal cross section (in fb) and diphoton invariant mass 
distribution (in fb/GeV) for the backgrounds after the cuts of 
Eqs.~(\ref{eq:1},\ref{eq:2},\ref{eq:3}). The squares are the Higgs signal for
$m_H = 100,110,120,130,140,150$~GeV.  The solid line represents the sum of all
backgrounds, with individual components from the irreducible QCD background 
(dashed line), the irreducible EW background (dot-dashed line), and for the 
double parton scattering (DPS) background (dotted line) shown below. 
\label{fig:Mgamgam}
}
\vspace*{0.2in}
\end{figure}

Fig.~\ref{fig:Mgamgam} shows the results after the cuts of Eq.~(\ref{eq:3}).  
This plot compares the total signal cross section, in fb, to the di-photon
invariant mass distribution, $d\sigma/dm_{\gamma\gamma}$ in fb/GeV and thus 
indicates the relative size of signal and background for a mass resolution
of $\Delta m_{\gamma\gamma}\approx\pm 0.5$~GeV. Actual resolutions are 
expected to be $\approx \pm 0.6\cdots 1$~GeV for CMS and about $\pm 1.5$~GeV
for ATLAS~\cite{CMS-ATLAS}. For our cuts, with 10~fb$^{-1}$ of data, we thus 
expect 13 to 21 $H \rightarrow \gamma \gamma$ events on a background of
14 to 7 events (for a resolution of $\pm 1$~GeV).  This corresponds to a
3.5 to 6.9 standard deviation signal. Thus, a Higgs boson discovery with a
mere 10~fb$^{-1}$ of data appears feasible in the $qq\to qqH\to jj\gamma\gamma$
channel. It should be noted, however, that this estimate does not consider
detector efficiencies, nor a detailed analysis of resolution effects. Such a
more detailed analysis is needed because more than 50\%
of the signal events have at least one jet with $|\eta| \leq 2.4$ (see
Fig~\ref{fig:y_dist}), leading to charged particle tracks in the central
detector.  As a result, the position of the interaction vertex can be more
accurately obtained, leading to improved photon invariant mass resolution.
We leave detailed studies of detector performance to the experimental
collaborations.

Limited detector efficiencies and resolutions can be compensated by exploiting
another feature of the $qq\to qqH$ signal, namely the absence of color exchange
between the two scattering quarks. As has been demonstrated for the analogous
$qq\to qqZ$ process, with its very similar kinematics~\cite{rsz}, $t$-channel
color singlet exchange leads to soft jet emission mainly in the very forward 
and very backward regions, and even here mini-jet emission is substantially 
suppressed compared to QCD backgrounds. QCD processes are dominated by 
$t$-channel color octet exchange which results in minijet emission mainly
in the central detector. These differences can be exploited in a central
minijet veto~\cite{bpz_minijet} in double forward jet tagging events. From 
previous studies of weak boson scattering signals~\cite{rsz,iz}, we expect 
a veto on additional central jets of $p_{Tj}\agt 20$~GeV to further reduce
the QCD and DPS backgrounds by up to one order of magnitude, while affecting
the signal at only the 10-20\% level.  These issues will be studied for the
$H\to \gamma\gamma$ signal also~\cite{rzgap}. With this additional background 
suppression we expect the discovery of the Higgs boson in the 
$qq\to qqH \to jj\gamma\gamma$ channel to be largely background free,
and possible with an integrated luminosity of 10~fb$^{-1}$ even when taking 
into account reduced detector efficiencies~\cite{CMS-ATLAS}.

%
%

\acknowledgements
This research was supported in part by the
University of Wisconsin Research Committee with funds granted by the
Wisconsin Alumni Research Foundation and in part by the U.~S.~Department
of Energy under Contract No.~DE-FG02-95ER40896.

%
%

\end{document}